\documentclass[aps,prl,superscriptaddress,amsfonts,amsmath,amssymb,showpacs,floatfix,reprint,longbibliography]{revtex4-2}
\usepackage{txfonts}
\usepackage{graphicx}
\usepackage{xcolor}

\newcommand{\css}{Co$_3$Sn$_2$S$_2$}

\begin{document}

\title{Doping Dependence of the in-Plane Transition in \css}

\author{Ivica {\v Z}ivkovi{\' c}*}
\email{ivica.zivkovic@epfl.ch}
\affiliation{Laboratory for Quantum Magnetism, Institute of Physics, \'Ecole Polytechnique F\'ed\'erale de Lausanne, CH-1015 Lausanne, Switzerland}

\author{Mohamed A. Kassem}
\affiliation{Department of Materials Science and Engineering, Kyoto University, Kyoto 606-8501, Japan}
\affiliation{Department of Physics, Faculty of Science, Assiut University, 71516 Assiut, Egypt}

\author{Yoshikazu Tabata}
\affiliation{Department of Materials Science and Engineering, Kyoto University, Kyoto 606-8501, Japan}

\author{Takeshi Waki}
\affiliation{Department of Materials Science and Engineering, Kyoto University, Kyoto 606-8501, Japan}

\author{Hiroyuki Nakamura}
\affiliation{Department of Materials Science and Engineering, Kyoto University, Kyoto 606-8501, Japan}

\date{\today}

\begin{abstract}
In \css two transitions are observed, the main one to a ferromagnetic state at $T_C = 174$\,K and the second one, involving in-plane components at $T_P = 127$\.K. We follow their doping dependence as Sn is replaced with In, which causes a reduction of $T_C$ and $T_P$. Importantly, both transitions follow the same doping dependence, indicating a single energy scale involved with both processes.
\end{abstract}

\maketitle

\css~is a candidate Weyl semimetal material, shown to exhibit anomalous Hall effect~\cite{Liu2018} and surface Fermi arcs~\cite{Liu2019}. Weyl semimetals are characterized by the existence of Weyl nodes, acting as sources and sinks of Berry curvature~\cite{Nagaosa2010}, which when found close to the Fermi energy are contributing to anomalous transport properties. Their number and the location in the \textit{k}-space are intimately linked to the material's magnetic order. In \css\,it has been demonstrated that the separation of Weyl nodes is linked to the value of the magnetic moment~\cite{Wang2018}. On the other hand, applying the magnetic field perpendicular to the hexagonal \textit{c}-axis shifts, creates and annihilates Weyl nodes as the moment is canted towards the kagome plane~\cite{Ghimire2019}.

The properties of the magnetic order in \css\,have been under scrutiny for some time. Although established as an itinerant ferromagnet~\cite{Schnelle2013}, successive reports have indicated a more complicated picture emerging below $T_C = 174$\,K. Antiferromagnetic~\cite{Guguchia2020} and spin-glass~\cite{Lachman2020} phases have been invoked, with later reports showing inconsistencies with these proposals~\cite{Soh2022,Zivkovic2022}. The existence of an additional transition has been proposed~\cite{Kassem2017}, with a detailed angular dependence showing that it is linked with the in-plane projection of moments, which are slightly tilted away from the \textit{c}-axis and form an umbrella structure~\cite{Zivkovic2022}.

In this short note we present experimental evidence of the intrinsic origin of the in-plane transition in \css\,revealed by its dependence on In-doping. By replacing Sn with In the number of electrons is reduced which leads to a suppression of the density of states at the Fermi level and a subsequent reduction of the main ordering temperature. Previous studies on single crystals~\cite{Kassem2015} have pointed out that the critical concentration is around 0.8, with larger doping levels resulting in a non-magnetic ground state. Here we extend those studies by focusing also on the second transition recently revealed to involve in-plane components of magnetic moments~\cite{Zivkovic2022}.

The preparation of single crystals has been described previously~\cite{Kassem2015}. Samples were prepared by flux and Bridgman methods and their precise composition has been determined by wave length dispersive X-ray spectroscopy (WDX). We did not observe any systematic difference in the properties of samples related to the growth method. The specimens of an appropriate size were mounted on a sample holder which permits the rotation around an axis perpendicular to the direction of magnetic field $B$ with a resolution better than 0.1$^0$. Both ac and dc signals were measured on a superconducting quantum interference device (SQUID)-based magnetometer (MPMS3, Quantum Design) with respect to temperature in zero-field-cooled (ZFC)/field-cooled (FC) regimes. Zero-field state has been ensured by resetting the superconducting magnet.


\begin{figure}[b]
\centering
\includegraphics[width=0.95\columnwidth]{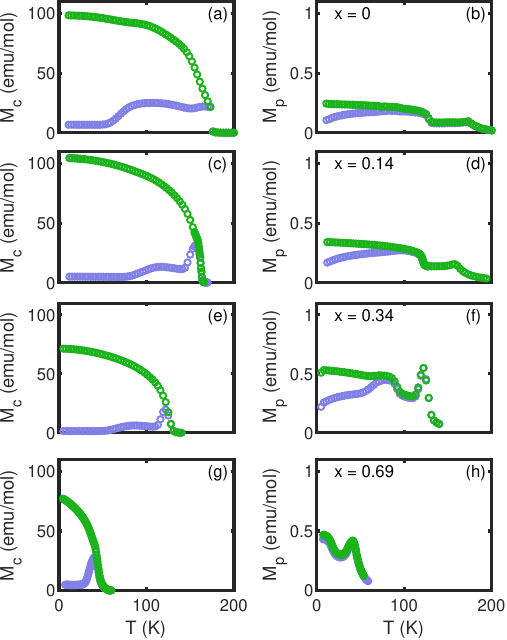}
\caption{(Color online) Magnetization of \css\,parallel to ((a), (c), (e) and (g)) and perpendicular to the \textit{c}-axis ((b), (d), (f) and (h)) measured in $B = 1$\,mT.}
\label{fig:magnetization}
\end{figure}

ZFC/FC temperature scans were measured with magnetic field $B = 1$\,mT parallel and perpendicular to the \textit{c}-axis. Figure~\ref{fig:magnetization} displays the evolution of the magnetization with respect to the doping level of indium. The parent compound ($x = 0$) exhibits the main transition at $T_C = 174$\,K and the second transition at $T_P = 127$\,K, in accordance with previous results~\cite{Zivkovic2022}. Both transitions are characterized by a sharp increase of magnetization and a splitting of ZFC and FC measurements. With In doping those features are shifted to lower temperatures, reflecting the decrease of electron density of states. At larger doping levels the splitting below $T_P$ for $B \perp c$ is significantly decreased, which we attribute to the decrease of energy barriers pinning the in-plane moment projections. One should note that the splitting of the in-plane component remains more than two orders of magnitude smaller compared to the \textit{c}-axis splitting across the doping levels.

\begin{figure}
\centering
\includegraphics[width=0.8\columnwidth]{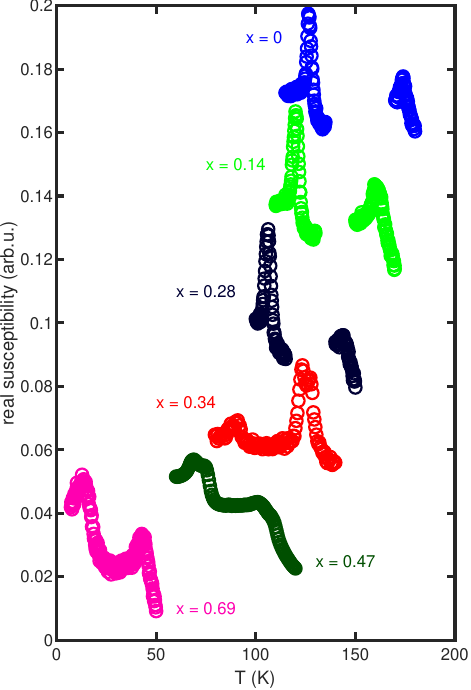}
\caption{(Color online) Real part of ac susceptibility across two transitions for several doping levels measured with alternating magnetic field perpendicular to the \textit{c}-axis.}
\label{fig:AC}
\end{figure}

Alongside the ZFC/FC splitting, each transition is accompanied by a sharp peak in ac magnetic susceptibility. In Figure~\ref{fig:AC} we show the real component of the susceptibility across transitions for four doping levels grown by the flux method ($x=0, 0.14, 0.34, 0.69$, their M(T) curves are presented in Figure~\ref{fig:magnetization}) and two doping levels grown by the Bridgman method ($x=0.28, 0.47$). Magnetization curves from Bridgman-grown samples are qualitatively the same as those of flux-grown samples.


\begin{figure}
\centering
\includegraphics[width=0.9\columnwidth]{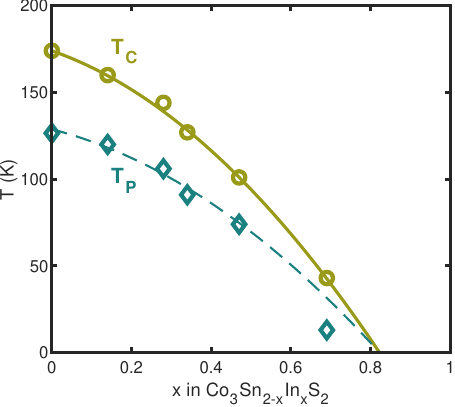}
\caption{(Color online) Doping dependence of the main transition $T_C$ and the in-plane transition $T_P$.}
\label{fig:diagram}
\end{figure}

We can now extract the temperatures of peaks and construct a phase diagram. In Figure~\ref{fig:diagram} dark circles represent the main transition $T_C$ while light diamonds stand for the in-plane transition at $T_P$. The solid line that follows the doping dependence of the main transition is a simple second-order polynomial which hints that around $x \sim 0.8$ a quantum phase transition occurs between the itinerant ferromagnetic phase and a nonmagnetic metal. This is in excellent agreement with previous findings~\cite{Kassem2015} where for $x > 0.8$ it has been indeed shown that the ground state is nonmagnetic.

If we now renormalize the second order polynomial on the temperature scale given by $T_P$, we obtain the dashed line in Figure~\ref{fig:diagram}. It is found that there is a very good agreement with experimental points for a wide doping level. It is only the highest doping level that deviates somewhat, which can potentially be assigned to a disorder effect on a fragile state of in-plane components.

The successful description of $T_P$ vs $x$ with a simple rescaling of $T_C$ vs $x$ dependence indicates that a single energy scale is responsible for both features. Most importantly, it proves that the feature observed at $T_P$ and the array of properties characterizing it, like strictly in-plane dynamics and orders of magnitude weaker signal, are intrinsic to \css.


\begin{thebibliography}{0}%
\makeatletter
\providecommand \@ifxundefined [1]{%
 \@ifx{#1\undefined}
}%
\providecommand \@ifnum [1]{%
 \ifnum #1\expandafter \@firstoftwo
 \else \expandafter \@secondoftwo
 \fi
}%
\providecommand \@ifx [1]{%
 \ifx #1\expandafter \@firstoftwo
 \else \expandafter \@secondoftwo
 \fi
}%
\providecommand \natexlab [1]{#1}%
\providecommand \enquote  [1]{``#1''}%
\providecommand \bibnamefont  [1]{#1}%
\providecommand \bibfnamefont [1]{#1}%
\providecommand \citenamefont [1]{#1}%
\providecommand \href@noop [0]{\@secondoftwo}%
\providecommand \href [0]{\begingroup \@sanitize@url \@href}%
\providecommand \@href[1]{\@@startlink{#1}\@@href}%
\providecommand \@@href[1]{\endgroup#1\@@endlink}%
\providecommand \@sanitize@url [0]{\catcode `\\12\catcode `\$12\catcode
  `\&12\catcode `\#12\catcode `\^12\catcode `\_12\catcode `\%12\relax}%
\providecommand \@@startlink[1]{}%
\providecommand \@@endlink[0]{}%
\providecommand \url  [0]{\begingroup\@sanitize@url \@url }%
\providecommand \@url [1]{\endgroup\@href {#1}{\urlprefix }}%
\providecommand \urlprefix  [0]{URL }%
\providecommand \Eprint [0]{\href }%
\providecommand \doibase [0]{https://doi.org/}%
\providecommand \selectlanguage [0]{\@gobble}%
\providecommand \bibinfo  [0]{\@secondoftwo}%
\providecommand \bibfield  [0]{\@secondoftwo}%
\providecommand \translation [1]{[#1]}%
\providecommand \BibitemOpen [0]{}%
\providecommand \bibitemStop [0]{}%
\providecommand \bibitemNoStop [0]{.\EOS\space}%
\providecommand \EOS [0]{\spacefactor3000\relax}%
\providecommand \BibitemShut  [1]{\csname bibitem#1\endcsname}%
\let\auto@bib@innerbib\@empty
\end{thebibliography}%


\begin{thebibliography}{9}
\bibitem{Liu2018} E. Liu, Y. Sun, N. Kumar, L. Muechler, A. Sun, L. Jiao, S.-Y. Yang, D. Liu, A. Liang, Q. Xu, J. Kroder, V. Süß, H. Borrmann,C. Shekhar, Z. Wang, C. Xi, W. Wang, W. Schnelle, S. Wirth, Y. Chen, S.T.B. Goennenwein, and C. Felser, Nat.Phys. 14, 1125 (2018)
\bibitem{Liu2019} D. F. Liu, A. J. Liang, E. K. Liu, Q. N. Xu, Y. W. Li, C. Chen, D. Pei, W. J. Shi, S. K. Mo, P. Dudin, T. Kim, C. Cacho, G. Li, Y. Sun, L. X. Yang, Z. K. Liu, S. S. P. Parkin, C. Felser, and Y. L. Chen, Science 365, 1282 (2019)
\bibitem{Nagaosa2010} N. Nagaosa, J. Sinova, S. Onoda, A.H. MacDonald, and N. P. Ong, Rev.Mod.Phys. 82, 1539 (2010)
\bibitem{Wang2018} Q. Wang, Y. Xu, R. Lou, Z. Liu, M. Li, Y. Huang, D. Shen, H.
Weng, S. Wang, and H. Lei, Nat.Commun. 9, 3681 (2018)
\bibitem{Ghimire2019} M.P. Ghimire, J.I. Facio, J.-S. You, L. Ye, J.G. Checkelsky, S. Fang, E. Kaxiras, M. Richter, and J. van den Brink, Phys.Rev.Res. 1, 032044(R) (2019)
\bibitem{Schnelle2013} W. Schnelle, A. Leithe-Jasper, H. Rosner, F.M. Schappacher,
R. Pottgen, F. Pielnhofer, and R. Weihrich, Phys.Rev.B 88, 144404 (2013)
\bibitem{Guguchia2020} Z. Guguchia, J.A.T. Verezhak, D.J. Gawryluk, S.S. Tsirkin, J.-X. Yin, I. Belopolski, H. Zhou, G. Simutis, S.-S. Zhang, T.A. Cochran, G. Chang, E. Pomjakushina, L. Keller, Z. Skrzeczkowska, Q. Wang, H.C. Lei, R. Khasanov, A. Amato, S. Jia, T. Neupert, H. Luetkens, and M.Z. Hasan, Nat.Commun. 11, 559 (2020)
\bibitem{Lachman2020} E. Lachman, R.A. Murphy, N. Maksimovic, R. Kealhofer,
S. Haley, R.D. McDonald, J.R. Long, and J.G. Analytis, Nat.Commun. 11, 560 (2020)
\bibitem{Soh2022} J.-R. Soh, C.J. Yi, I. Zivkovic, N. Qureshi, A. Stunault, B. Ouladdiaf, J.A. Rodriguez-Velamazan, Y.G. Shi, H.M. Ronnow, and A.T. Boothroyd, Phys. Rev. B 105, 094435 (2022)
\bibitem{Zivkovic2022} I. \v{Z}ivkovi\'c, R. Yadav, J.-R. Soh, C. Yi, Y. Shi, O. V. Yazyev, and H. M. R{\o}nnow, Phys. Rev. B 106, L180403 (2022)
\bibitem{Kassem2017} M. A. Kassem, Y. Tabata, T. Waki, and H. Nakamura, Phys.Rev.B 96, 014429 (2017)
\bibitem{Kassem2015} M.A. Kassem, Y. Tabata, T. Waki, and H. Nakamura, J. Crystal Growth 426, 208 (2015)
\end{thebibliography}

\end{document}